\documentclass[11pt]{article}
\usepackage{geometry}                
\usepackage{graphicx}
\usepackage{amssymb}
\usepackage{amsmath}
\usepackage{amsfonts}
\usepackage{graphicx}
\usepackage{epstopdf}
\usepackage{color}
\usepackage{cite}
\usepackage{fancyhdr}

\newcommand{\be}{\begin{equation}}
\newcommand{\ee}{\end{equation}}
\newcommand{\bea}{\begin{eqnarray}}
\newcommand{\eea}{\end{eqnarray}}
\newcommand{\beas}{\begin{eqnarray*}}
\newcommand{\eeas}{\end{eqnarray*}}

\title{How interbank lending amplifies overlapping portfolio contagion:  A case study of the Austrian banking network}
\date{\today}
\author{Fabio Caccioli$^{1,2}$, J. Doyne Farmer$^{1,2}$, Nick Foti$^{3}$, and Daniel Rockmore$^{1,3,4}$
~~~~~\\
{\em 1 - Santa Fe Institute, 1399 Hyde Park road, Santa Fe, NM 87501, USA}\\
{\em 2 - Institute for New Economic Thinking at the Oxford Martin School and Mathematical}\\
{{\em Institute}, Oxford University, Eagle House Walton Well Road Oxford OX2 6ED}\\ 
{\em 3 - Department of Computer Science, Dartmouth College, Hanover, NH 03755}\\
{\em 4 - Department of Mathematics, Dartmouth College, Hanover, NH 03755}\\
}

\begin{document}
\maketitle
\thispagestyle{fancy}

\begin{abstract}
In spite of the growing theoretical literature on cascades of failures in interbank lending networks, empirical results seem to suggest that networks of direct exposures are not the major channel of financial contagion.
In this paper we show that networks of interbank exposures can however significantly amplify contagion due to overlapping portfolios. To illustrate this point, we consider the case of the Austrian interbank network and perform stress tests on it according to different protocols. We consider in particular contagion due to (i) counterparty loss; (ii) roll-over risk; and (iii) overlapping portfolios. We find that the average number of bankruptcies caused by counterparty loss and roll-over risk is fairly small if these contagion mechanisms are considered in isolation. Once portfolio overlaps are also accounted for, however, we observe that the network of direct interbank exposures significantly contributes to systemic risk.

\end{abstract}

\tableofcontents
\section{Introduction}
The economic and financial crises of the early twenty--first century give strong indications that the high level of interconnectivity characterizing much of the contemporary economic system can amplify and propagate the stress originated in a specific economic sector or a specific financial institution to other sectors and other institutions \cite{Babus09}.

Connections between financial institutions are of various kinds, ranging from common assets held in balance sheets of different institutions to direct linkages between institutions corresponding to specific transactions. While such connectivity can serve as a means of risk management or increased efficiency for these institutions, it can also provide 
channels for contagion, thereby creating potential sources of systemic risk.
It is for this reason that understanding the nature and structure of connections between financial institutions and their impact on the system as a whole is of primary importance for the assessment of robustness of financial systems.

In ref. \cite{Upper}, Upper discusses and categorizes much of the recent work on robustness and stability in financial markets.  The basic object of study is a network of banks where edges encode financial exposure.  Most of these papers (see for instance \cite{Staum12})
model shocks to the system using variants of an algorithm developed by Furfine \cite{Furfine}.  The three steps of this iterative algorithm are to 1) create an extinction of a single bank, 2) calculate consequent extinctions via exposure to the original bank above an equity threshold, and 3) spread the contagion to other banks that have exposure to the now extinct banks.  Thus, Furfine's algorithm  also provides a method for assessing risk associated with counterparty loss.   

This basic threshold extinction model has been modified in various ways to include safety nets  \cite{upperworms2004}, risk management strategies \cite{elsinger2006}, different illiquidity conditions \cite{Furfine}, probabilities of contagion (vs. thresholds) \cite{Frisell,SM1998,Muller}, and  the integration of the topology of the banking network \cite{allen,Nier,battiston,drehmann-tarashev}. 
Of particular relevance is  M\"uller's detailed analysis of the Swiss banking network \cite{Muller}, where in addition to interbank exposures, the risk associated with the existence of credit lines between institutions is taken into account. 

The main question we address in this paper is if networks of direct interbank exposures play any role as an amplification mechanism for financial contagion. Empirical studies as well as recent theoretical developments seem to suggest that, in realistic scenarios, networks of direct exposures are not important sources of systemic risk \cite{Young13,Upper}; instead the main contribution comes from common asset holding\cite{Huang13,Caccioli12,Corsi13}. The point we make in this paper is that although the network of interbank lending may not by itself trigger global cascades of failures in the banking system, it can in certain regimes amplify the stress caused by common asset holdings. This indicates that the interaction between different contagion channels is important, creating risks that may be much larger than any single channel of contagion alone.

In this paper we focus on the Austrian interbank network, for which we perform stress tests according to different protocols. We also characterize the statistical properties of the balance sheets in this system and of the network of mutual exposures. The Austrian banking system has been previously studied \cite{AUSTRIA,AUSTRIA2}. This work differs from that not only in the time period of study (which allows us to do some comparison of statistics over the different epochs), but more importantly, in that our primary focus is on the analysis of the various contagion pathways. The national nature of the data also places our work in the growing body of literature and results that are now being generated for individual national banking systems (e.g., \cite{Brazil,Italy,Muller}). Collectively, this kind of work can ultimately enable a lessons-learned approach to the study of network aspects of the stability of interbank systems as well as the role of regulation in this setting.

Our empirical investigation draws heavily on a growing corpus of theoretical and modeling efforts that aim to  quantify the relation between network topology and contagion effects (see for instance\cite{allen,battiston,Nier,drehmann-tarashev,Gai11}). 
These kinds of network stress tests have their analogs in other complex systems, most notably, the {\em in silico} ``knockout" experiments or ``extinction analyses" that have been performed in a variety of network contexts~\cite{Albert00} including metabolic networks \cite{Jeong2000}, protein networks \cite{Jeong2001},    food web models of ecosystems (see e.g., Section 4.6 of \cite{dunne2006} and the many references therein as well as  \cite{Allesina2009}) and most recently, even the macroeconomic system that is the World Trade Web (WTW) \cite{fpr}. Of some relevance is the ``robust yet fragile" (RYF) categorization of networks. This is shorthand for networks that are resilient (according to some basic statistic such as diameter) under the (suitably defined) failure  of a random node, but will experience a rapid degradation (in terms of the measured statistic) under a judicious targeting of nodes for failure. This has been shown to be related to power law degree distributions \cite{pnas1,sf1,sf2,Albert00} and/or small world characteristics \cite{sw1,sw2,Xia}. In the case of the WTW, path length-related measures of robustness do not seem to be appropriate and the RYF characterization is  generalized accordingly \cite{fpr}. 

The paper is organized as follows: In the next section we introduce the data-set considered here and provide a characterization of the statistical properties of balance sheets and of the topological properties of the interbank network. The properties of the Austrian banking system that we find are consistent with previous studies concerning other national interbank systems. This similarity suggests that the conclusions of our paper can potentially be extended to interbank systems other than the specific one here considered. In Section \ref{directExposures} we present results of stress tests where contagion is due to counterparty loss and roll-over risk. In Section \ref{overlappingPortfolios} we consider what happens when overlapping portfolios are also accounted for, and explicitly show that the network of direct interbank exposures can in certain regimes greatly amplify contagion due to overlapping portfolios. We present our conclusions in Section \ref{conclusions}.

\section{The Data}\label{data}
The data studied here\footnote{Our data was made available by the Oesterreichische Nationalbank. We would like to thank
Claus Puhr and Martin Summer for their help in sharing and processing the data.}  contain information on the balance sheets  of Austrian banks. Data are available on a quarterly basis for the years 2006, 2007 and 2008 and consist of the following:

\begin{itemize}
\item  interbank claims encoded in an exposure matrix $L$, whose entries $L_{ij}$ represent  the liabilities bank $i$ has towards bank $j$\footnote{Reporting data up until year-end 2007 excluded short-term interbank lending with a maturity of less than one month.};
\item total liabilities (including non-interbank liabilities) $L_i^{tot}$; 
\item total assets (including non-interbank assets) $A_i^{tot}$;
\item total amount of liquid assets $A_i^{liq}$, i.e., assets that can be easily liquidated;\footnote{Data provided by the Oesterreichische Nationalbank includes only highly liquid assets, i.e. cash and reserves held with the central bank.}
\end{itemize}

The number of banks for which information is provided changes slightly from quarter to quarter as detailed in the following table:\\
 \begin{center}
 \begin{tabular}{ |c|| c| c| c|c|}
 \hline
 & $1^{st}$ quarter & $2^{nd}$ quarter & $3^{rd}$ quarter & $4^{th}$ quarter\\
 \hline 
 \hline
 2006 & 846 & 844 & 832 & 832 \\
  \hline
  2007 & 834 & 834 & 825 & 825\\
  \hline
  2008 & 825 & 828 & 824 & 832\\
  \hline
\end{tabular}
\end{center}

A similar data-set for the Austrian banking system concerning the period 2003-2006 has been previously studied in \cite{AUSTRIA}.

Unless otherwise stated, we consider in the following subsections results obtained by aggregating data from all the quarters at our disposal. We have performed analysis also for subsets corresponding to single quarters and seen that the properties of the system are stable over time and consistent with that observed at the aggregate level.

\subsection{Statistical properties of balance sheets}

\begin{center}
\begin{figure}[h]
\begin{center}
\includegraphics[width=8cm]{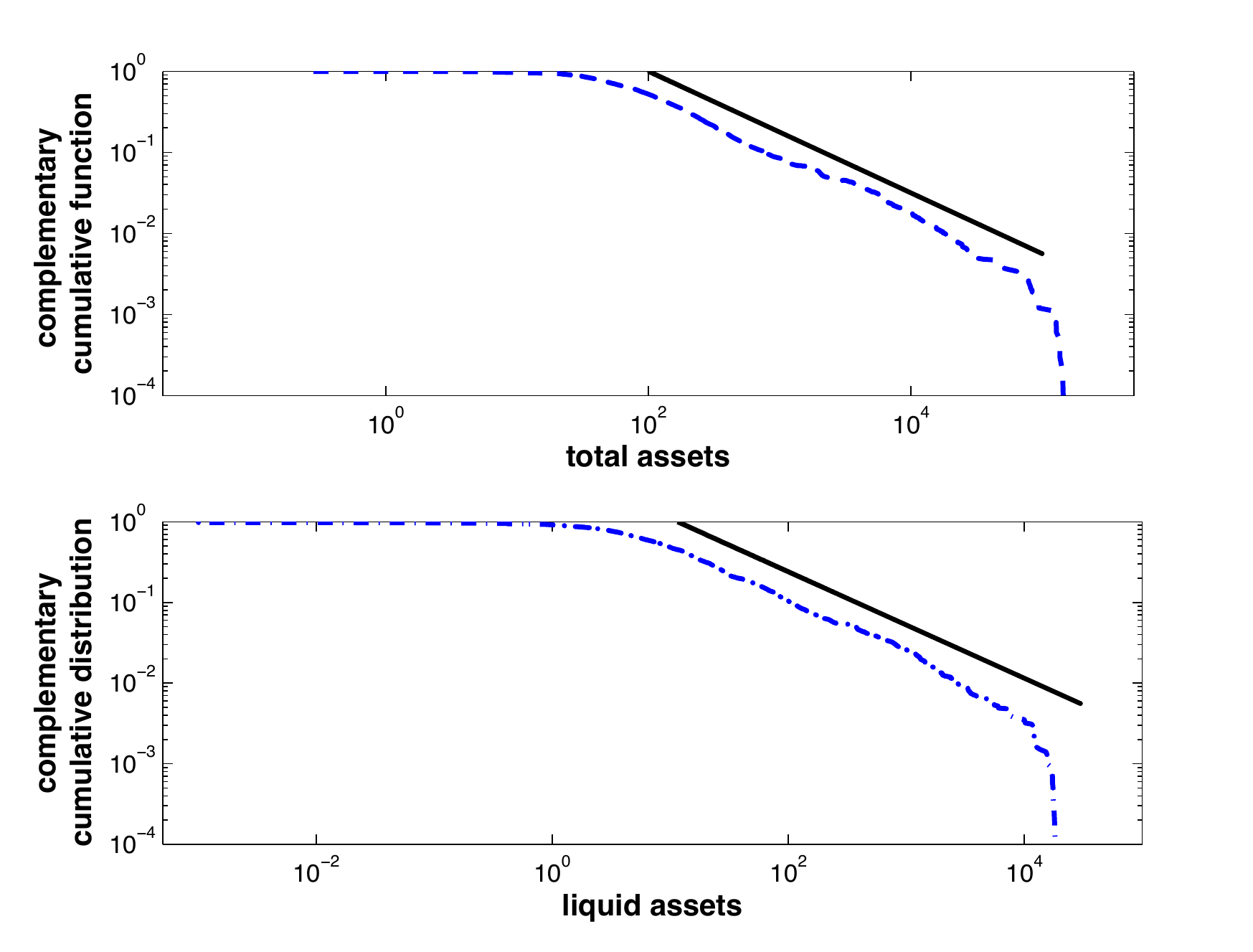}
\caption{\footnotesize {\textit{Statistics of banks balance sheet. {\bf Upper panel}: Complementary cumulative distribution of total assets (blue dashed line). Data are aggregated from all $12$ quarters and are presented on a log-log scale. We observe three regimes: {\em i)} banks with balance sheet size smaller than $60$ M euro corresponding to the flat part of the distribution; {\em ii)} typical banks with a balance sheet size between approximately $60$ M and $35$ B, for which the distribution can be approximate by a power law behavior; {\em iii)} big banks with balance sheet size greater than $35$ B, that characterize the cutoff of the distribution. The black solid line represents a power law of exponent $0.74$ obtained from a least squares fit of banks in the ``typical'' regime.
{\bf Bottom panel}: Complementary cumulative distribution of liquid assets.  We observe again three regimes: {\em i)} banks with small amount of liquid assets (less than $1$ M) for which the distribution is flat; {\em ii)} banks with an intermediate amounts of liquid assets (between $1$ M and $1$ B) for which the distribution is approximated by a power law; {\em iii)} banks with large amounts of liquid assets (more than $1$ B) in the cutoff of the distribution.  
The black solid line represents a power law of exponent $0.67$ obtained from a least squares fit of banks in the ``typical'' regime.
}}}\label{assets}
\end{center}
\end{figure}
\end{center}

In Figure \ref{assets} we show complementary cumulative distributions\footnote{We define the complementary cumulative distribution of a probability density function $f(x)$ the quantiy $F(y)=\int_y^{\infty}dx f(x)$, i.e., the probability that a random variable distributed as $f(x)$ is greater than $y$.} for the amount of total assets, total liabilities and liquid assets held by banks in their balance sheets. The  three quantities display a similar pattern characterized by an initial regime with a flat distribution, a ``typical'' regime where the distribution roughly follows a power law distribution, and a subsequent  cut-off regime. 

As far as total assets and liabilities are concerned (top panel of Figure \ref{assets}), in the first regime we have $3253$ data-points with total assets smaller than (approximately) $60$ million euro, while in the cutoff regime $49$ data-points have total assets greater than (approximately) 35 billion euro. The remaining $6687$ data-points constitute the intermediate ``typical'' regime, that spans an extended range of values (approximately 3 orders of magnitude) and can be described in terms of a power law with exponent $\simeq 0.75$.
The bottom panel of Figure \ref{assets} shows a similar structure for the amount of liquid assets held by banks in their balance sheet. In this case the initial regime is made of banks with less than $5$ million euro of liquid assets, while in the cutoff regime  banks have more than $10$ billion euro of liquid assets. The intermediate regime (ranging over three orders of magnitude) can again be characterized in terms of a power law, with an exponent $\simeq 0.67$.
Note that banks in the ``typical'' regime for total assets and liabilities are not necessarily in the ``typical'' regime for liquid assets (and vice versa).
\begin{center}
\begin{figure}[h]
\begin{center}
\includegraphics[width=10cm]{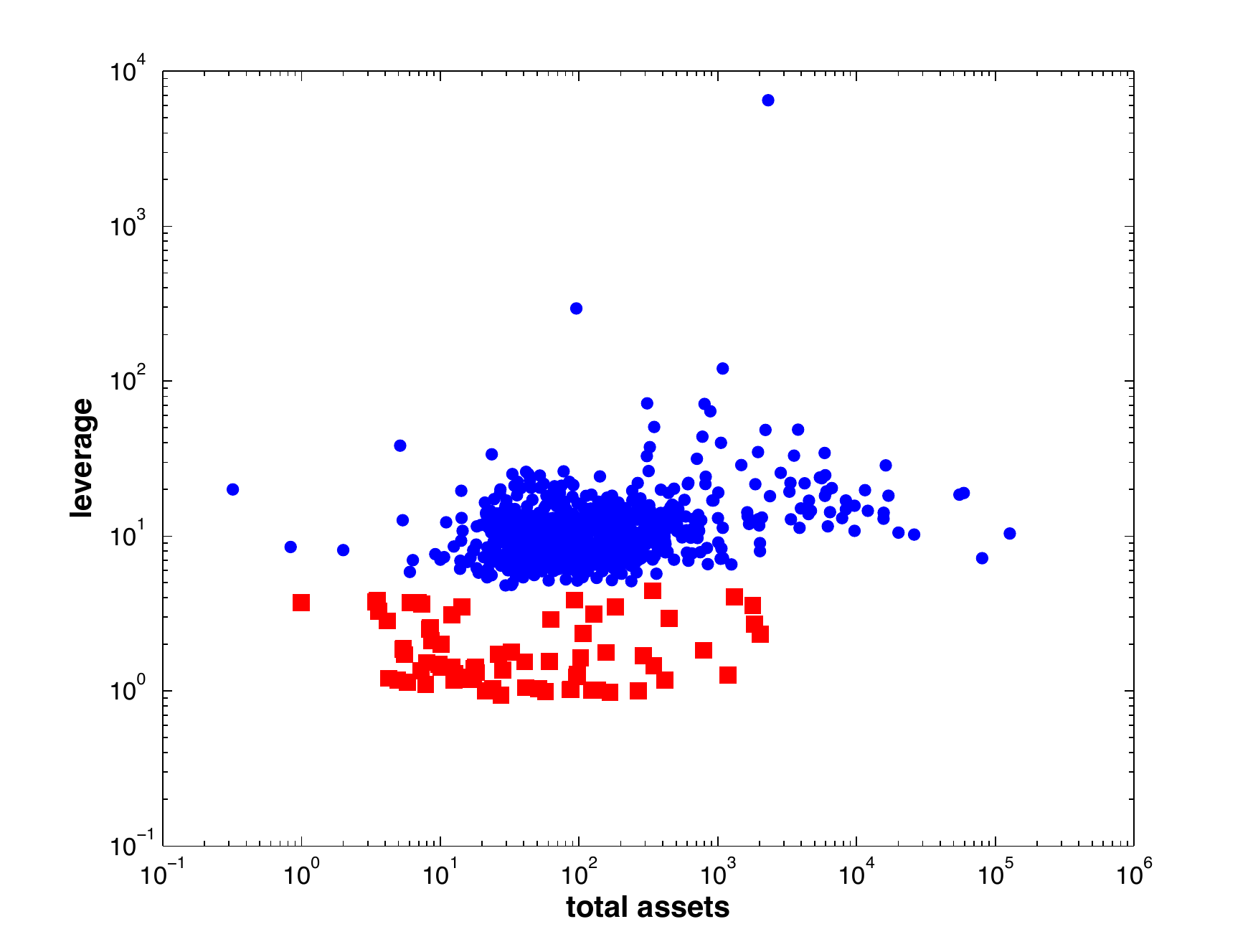}
\caption{\footnotesize {\textit{Scatter plot of leverage vs assets. Banks can be roughly clustered into two different groups: Red squares refer to banks with leverage smaller than $4.6$; the remaining institutions are represented by blue dots.  Data refer to the first quarter of 2006.}}}\label{figure2}
\end{center}
\end{figure}
\end{center}

 \begin{center}
\begin{figure}[h]
\begin{center}
\includegraphics[width=10cm]{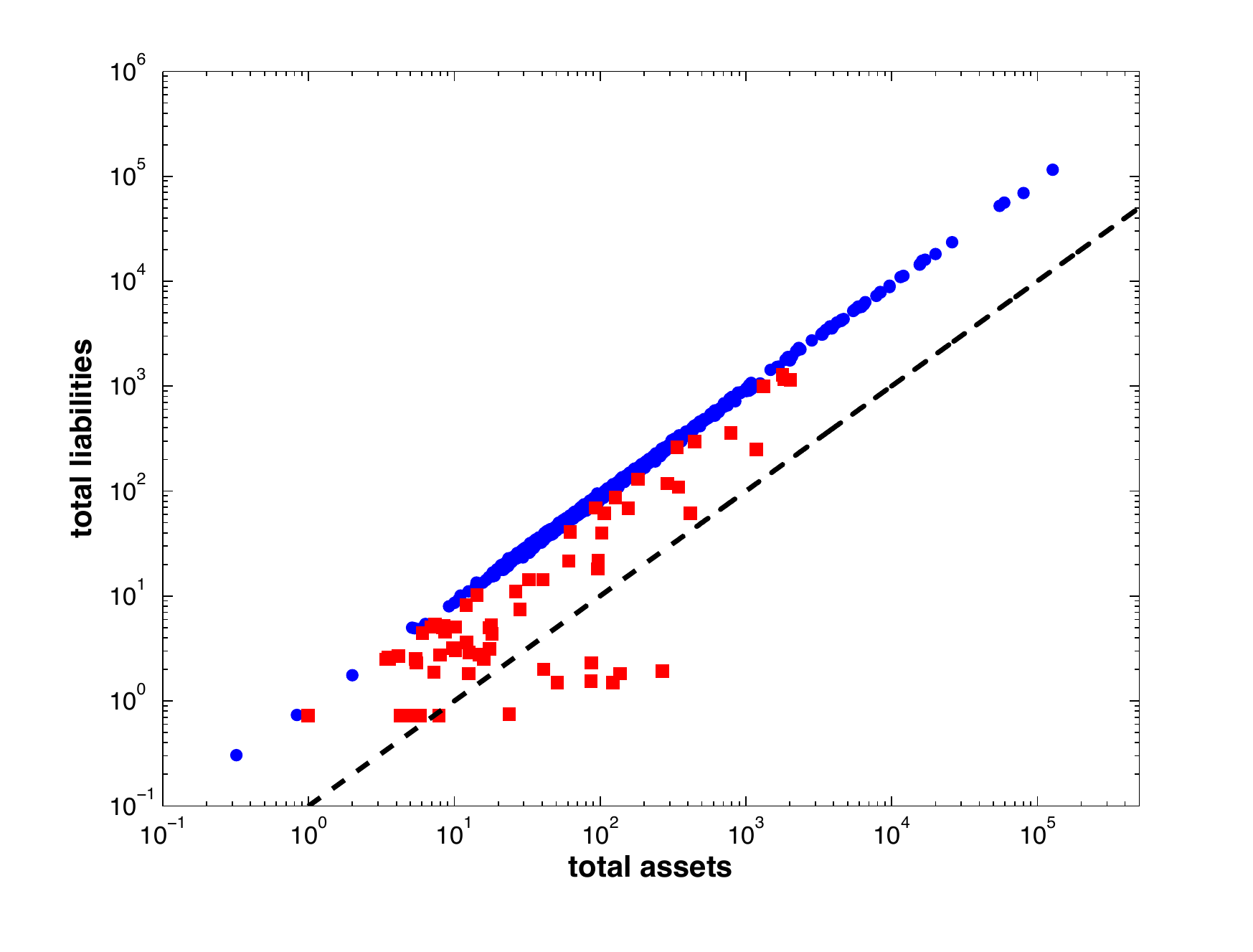}
\caption{\footnotesize {\textit{Scatter plot of total assets vs liabilities in log-log scale. Different symbols and colors correspond to the same grouping of Figure \ref{figure2}, and also seem to characterize the system. 
Banks with leverage higher than $4.6$ seem to share the same linear relation between total assets and liabilities. Deviations from the linear pattern are observed for banks with leverage smaller than $4.6$. 
The black dashed line represents a line of slope $1$, that in the double logarithmic scale denotes a linear relation between total assets and liabilities.
 Data refer to the first quarter of 2006.}}}\label{figure3}
\end{center}
\end{figure}
\end{center}
An important parameter to characterize the investment strategy of financial institutions is the leverage, that is the ratio between total assets and equity. This measures the level of borrowing that banks use to finance their investments:
\be
\lambda_i=\frac{A_i^{tot}}{A_i^{tot}-L_i^{tot}}.
\ee

In Figure \ref{figure2} we present a scatter plot of total assets vs leverage relative to the first quarter of 2006\footnote{We limit our study to the first quarter of 2006 for convenience. Analysis of other quarters produces similar plots.}. From the plot, two different groups of banks seem to emerge:
\begin{itemize}
\item {\bf Region I}: a region made of $765$ banks with  leverage higher than $4.6$ (blue dots);
\item {\bf Region II}: a region at the bottom of the plot made by $71$ banks with leverage smaller than $4.6$.\footnote{{The data sample includes a number of special purpose banks (for instance pension funds or treasury departments of large corporates) as well as private banks. Neither of these engage in traditional lending business and should therefore have structurally lower leverage. We cannot however claim that these banks correspond to those observed in region II as we do not have bank identifiers.}}
\end{itemize}
The naive grouping emerging from Figure \ref{figure2} appears to be relevant to characterize the pattern emerging in Figure  \ref{figure3}, where we present a scatter plot of total assets vs total liabilities.

In Figure \ref{figure3} we present a scatter plot of total assets vs total liabilities.
We see that banks with leverage higher than $4.6$ (blue dots) are characterized by a clear linear relation between their total assets and their total liabilities.
Banks with leverage smaller than $4.6$ (red squares) represent instead a deviation from the overall linear pattern, characterized by the relation $L_i^{tot}\simeq 0.91A_i^{tot}$. The overall linear relation between total assets and liabilities suggests the presence of a typical value of leverage used by banks $\lambda_0\simeq 11$. In summary, we observe the existence of a typical leverage targeted by institutions with a quite heterogeneous pattern of balance sheet size \cite{Adrian09}.

\subsection{Topological properties of the network} 
We now turn to the characterization of the topological properties of the networks of direct exposures between banks. This characterization is important because results of complex network theory show that there is a relation between topology of networks and properties of dynamical processes taking place on them (see e.g. \cite{Barrat08}). Showing that the data-set considered in this study shares some of the properties previously observed for other interbank networks is then important as it suggests the generality of the contagion dynamics observed here.  

Given the exposure matrix $L_{ij}$, we can characterize the topological properties of the network  by introducing binary directed and undirected adjacency matrices for the network. Each of these matrices will use the notation $X$. A directed liability matrix $X^{lia}$ can be built in such a way that $X^{lia}_{ij}=1$ if bank $i$ borrowed money from bank $j$. This is just the exposure matrix thresholded at zero. 

A complementary asset matrix $X^{asset}$ can be built by assigning $X^{asset}_{ij}=1$ if bank $i$ lent money to bank $j$. Thus, ${\left(X^{asset}\right)}'=X^{lia}$ where the superscript prime denotes transpose. Finally, an undirected matrix $X$ can be built such that $X_{ij}=X_{ji}=1$ whenever a relation exists between banks $i$ and $j$, i.e., 
$$X_{ij} = \max{\{X^{lia}_{ij}, X^{asset}_{ij}\}}.$$ 

Previous empirical work has shown that interbank networks are characterized by heavy-tailed degree distributions and negative degree correlations  \cite{AUSTRIA,Italy,Belgium}, and theoretical work has also been carried on to understand their impact on contagion dynamics\cite{Caccioli12b,Georg13,Lenzu12}.  We find that the same topological properties apply to the data-set considered here.

\begin{center}
\begin{figure}[h]
\begin{center}
\includegraphics[width=8cm]{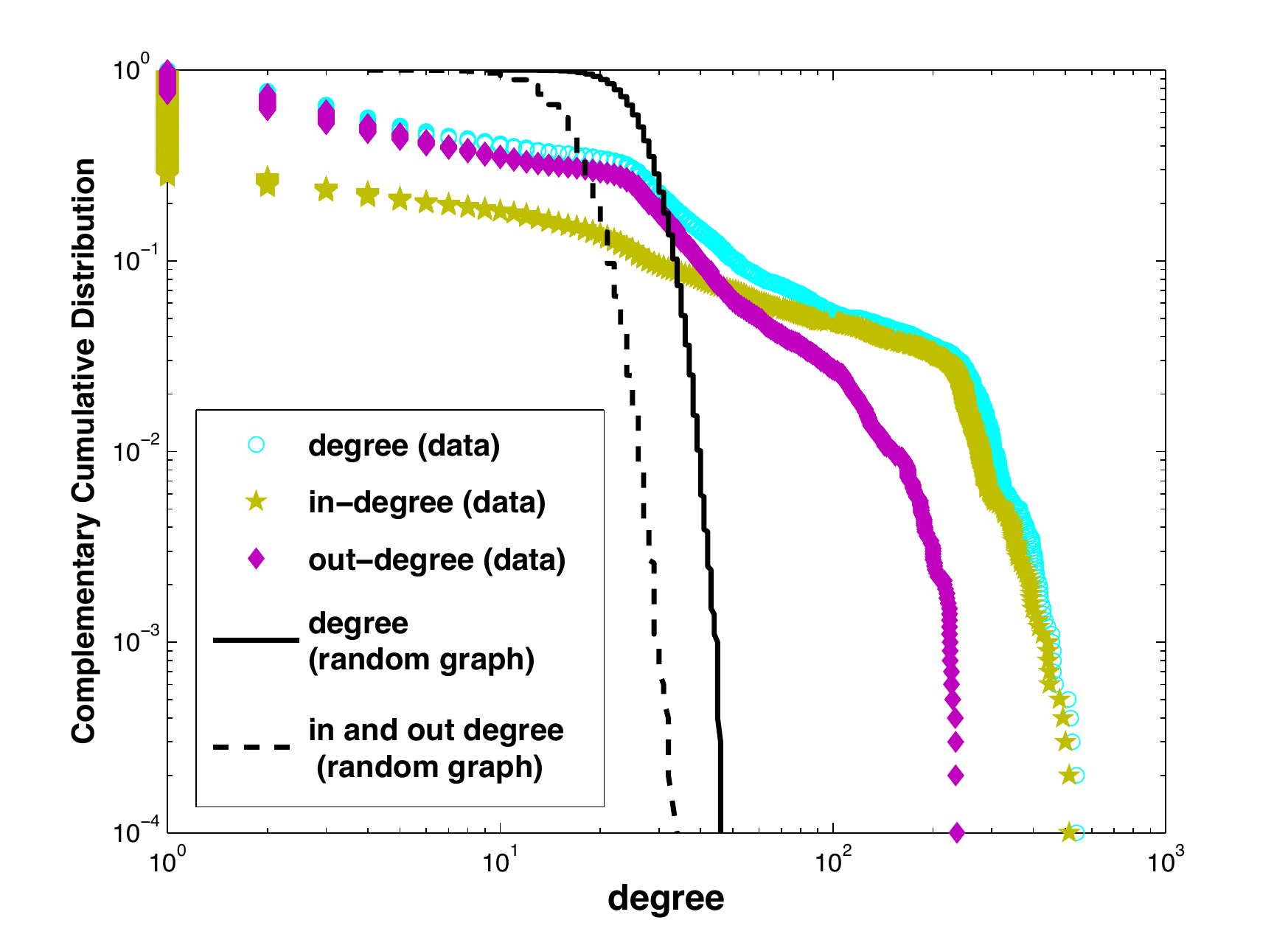}
\caption{\footnotesize {\textit{Degree distribution. Cyan circle: complementary cumulative distributions of node degrees, computed from the undirected adjacency matrix $X$ as $k_i=\sum_j X_{ij}$. Yellow stars: complementary probability distributions of node in-degrees, computed from the directed matrix $X^{asset}$ as $k^{in}_i=\sum_j X^{asset}_{ij}$. Magenta diamonds: complementary cumulative distributions of node out-degrees, computed from the directed matrix $X^{lia}$ as $k^{out}_i=\sum_j X^{lia}_{ij}$.  Black solid line: complementary cumulative distribution of node degrees for Erd\H{o}s-Renyi random graphs with the same average degree. Black dashed line: complementary cumulative distribution of node degrees for directed Erd\H{o}s-Renyi random graphs with the same  in/out average degree. Data are aggregated from all $12$ matrices. The degree distribution of the real system appears to be heavy-tailed.}}}\label{degree}
\end{center}
\end{figure}
\end{center}

In Figure \ref{degree} we show the complementary cumulative distributions of in-degrees ($\sum_j X^{asset}_{ij}$), out-degrees ($\sum_j X^{lia}_{ij}$) and degrees  ($\sum_j X_{ij}$). There is no obvious power law, but in all the three cases the distribution is characterized by a heavy tail, as 
can be observed by the comparison with the corresponding distributions obtained for Erd\H{o}s-Renyi random graphs\footnote{Erd\H{o}s-Renyi random graphs are constructed by fixing a probability $p$ of connecting any two nodes and performing the independent coin-tosses edge-by-edge. Note that in this case for a graph with $n$ nodes, the expected degree is $p(n-1)$, so given a graph with average degree $c$, the appropriate wiring parameter is $p=c/(n-1)$. Similar networks can be generated for the directed case by assigning random directions to links.}  with the same average degree. 
The presence of heavy tails in the degree distribution is usually associated with a higher overall robustness of the network with respect to the failure of a random node, but with a higher fragility with respect to ``targeted failures'' of highly connected nodes \cite{Albert00,Caccioli12b}\\

The level of correlations can be measured through the usual degree assortativity (cf. \cite{NewmanBook}) that quantifies the extent to which nodes of a given degree link to one another
\be
r=\frac{\langle k k'\rangle_l-\langle (k + k')/2\rangle_l^2}{\langle (k^2 + k'^2)/2\rangle_l-\langle (k + k')/2\rangle_l^2}.
\ee
Here $\langle\cdots\rangle_l$ denotes the average over all links and $k,k'$ the degrees of two nodes connected through a link.

We measured the assortativity for each of the $12$ undirected adjacency matrices, obtaining an average assortativity (averaged over the $12$ quarters) of $r_{av}\simeq - 0.62\pm 0.03$. The negative value indicates that in the interbank network  nodes of low degree tend to be connected with nodes of high degree, a structure characteristic of networks that are effectively of a ``hub and spoke" topology. 
As for the heavy-tailed nature of the degree distribution, the presence of correlations among degrees affects the probability of cascades triggered by the failure of a (random) bank \cite{Caccioli12} with respect to uncorrelated Erd\H{o}s-Renyi random graphs.

Clustering coefficients (see \cite{NewmanBook}) measure the tendency of neighbors of a given node to be linked to each other. The propensity of these networks to form triangles (directed or not) is a natural proxy for the financial interdependence of the institutions. The local clustering coefficient $C_i$  for a node $i$ is defined as
\be
C_i=\frac{\rm number~of~links~between~neighbors~of~i}{\rm number~of~possible~links~between~neighbors~of~i},
\ee
 and gives a measure of how the neighborhood of $i$ is close to be a clique. 
 
 The {\em average local clustering} is then measured as\footnote{Notice that the average clustering coefficient puts more weights on low connected nodes, and the observed high value may be driven by the hub and spoke structure of the network.} 
 $$\overline{C}= {1\over N} \sum_i C_i.$$
When we also average over the $12$ quarters at our disposal, we find that $\overline{C}=0.87\pm 0.02$, where the error has been computed as the standard deviation over the different quarters. This value can be compared to that of Erd\H{o}s-Renyi random networks of the same average degree, which have  $\overline{C}=0.032\pm 0.002$.

A comparison with a null model in which we randomly rewire links while preserving the in- and out- degrees of each of the nodes (often called the ``configuration model''), shows that the degree sequence is enough to reproduce both the negative correlations and the high local clustering observed in the data. There are however higher order topological properties that such null models cannot explain. For instance, the real networks appear to be characterized by a high number of directed loops when compared with synthetic networks (see appendix A).

\section{Stress tests: contagion through direct exposures}\label{directExposures}

In this section, we perform different stress tests on the interbank network to see to what extent failures of single banks propagate through the system.

{\bf Counterparty loss:}
The first  mechanism of contagion we consider is counterparty loss. This is related to the fact that if one institution goes bankrupt its creditors face a loss and can in turn go bankrupt if their equity is smaller than the loss. {Note in this respect that interbank liabilities are often bidirectional, meaning that bank $i$ borrows from bank $j$ as well as $j$ borrows from $i$. In the following we will consider for simplicity net exposures between banks, i.e. we define a matrix of net exposures ${L^{net}}$ with entries ${L}^{net}_{ij}=\max\{0,L_{ij}-L_{ji}\}$, where $L_{ij}$ is the amount of money that bank $i$ borrowed from $j$. The implicit assumption is that if $L_{ij}>L_{ji}$ and bank $i$ defaults on its obligations to $j$, $j$ will cover part of the loss by not paying its debt to $i$.
Let us define for convenience the capital (or equity) of bank $i$ as the difference between its total assets and liabilities: $C_i=A_i^{tot}-L_i^{tot}$, where $A_i^{tot}$ and $L_i^{tot}$ represent total assets and liabilities of bank $i$. For each bank, we also introduce the state variable $\sigma_i$ such that $\sigma_i=1$ if $i$ is bankrupt and zero otherwise.
The protocol used to probe the stability of the system with respect to counterparty loss is the following:
\begin{enumerate}
\item A seed node $q$ is selected and the corresponding bank is shut down. Here $\sigma_q=1$, while $\sigma_i=0$ for all other banks.
\item Banks revise their balance sheet. For each bank $i$ if the following condition holds then the bank fails:
\be
C_i<\sum_j L_{ji}^{net}\sigma_j,
\ee
where the right-hand side represent the loss of bank $i$, and we assume the extreme case where no money is recovered from failed banks. . If $i$ fails $\sigma_i$ is set to one.
\item If there are new bankruptcies return to step 2. Exit otherwise.
\end{enumerate}
}

{\bf Roll-over risk:}
A second contagion mechanism we consider is roll-over risk. This refers to the situation in which a bank that is used to borrowing money in the interbank market suddenly cannot raise the funds it needs to run its business. The recent economic crisis was characterized by the tendency of financial institutions to hoard liquidity resulting in a  corresponding freezing of the interbank market.  Here we perform a stress test to assess the probability that institutions  stop lending in the interbank market in response to a single bank hoarding liquidity\footnote{{The assumption that banks can stop lending in the interbank market is a strong one. In general, banks have contractual obligations and a bank could not, even if loans are not paid back. This is further amplified in the Austrian case due to the institutional ties of the multi-tiered Austrian banking sectors (savings and cooperative banks, where liquidity is often managed centrally).}}. {In this case, $\sigma_i=1$ if bank $i$ stops lending in the interbank market and zero otherwise
The stress test has been performed according to the following protocol:
\begin{enumerate}
\item A seed node $q$ is selected and the corresponding interbank assets are not rolled-over, that is $\sigma_q$ is set to one and $\sigma_i=0$ for all other banks. 
\item For all banks $i$, if the following condition holds bank $i$ stops lending in the interbank market and $\sigma_i$ is set to one:
\be
\sum_{j} L^{net}_{ij}\sigma_j>A^{liq}_i+f \left(A_i^{tot}-A_i^{liq}\right).
\ee
The left hand side of the above expression represents the shortage of funding that bank $i$ is facing due to other banks hoarding liquidity. Here we assume banks are able to liquidate on a short term a fraction $g$ ($0\le g\le1$) of their illiquid assets. 
\item If new banks stopped lending return to step 2. Exit otherwise.
\end{enumerate}
}
We define a {\em contagion event} as an event where {\bf at least} one bank goes bankrupt (counterparty risk) or starts hoarding liquidity (roll-over risk) as a response to the initial perturbation.

We performed these tests using every node of the network as a seed (for a comparison with similar results obtained for the configuration model, where links are randomly swapped, see Appendix B). Results are reported in Figure \ref{figure5} (blue dots) for the two observables: \\

\smallskip

\indent {\em i)}
{\it Contagion probability}, defined as the probability of observing a contagion event. \\
\indent  {\em ii)} {\it Conditional extent of contagion}, defined as the average fraction of banks affected by the initial shock if contagion occurs. 
\smallskip

\begin{center}
\begin{figure}[h]
\begin{center}
\includegraphics[width=8cm]{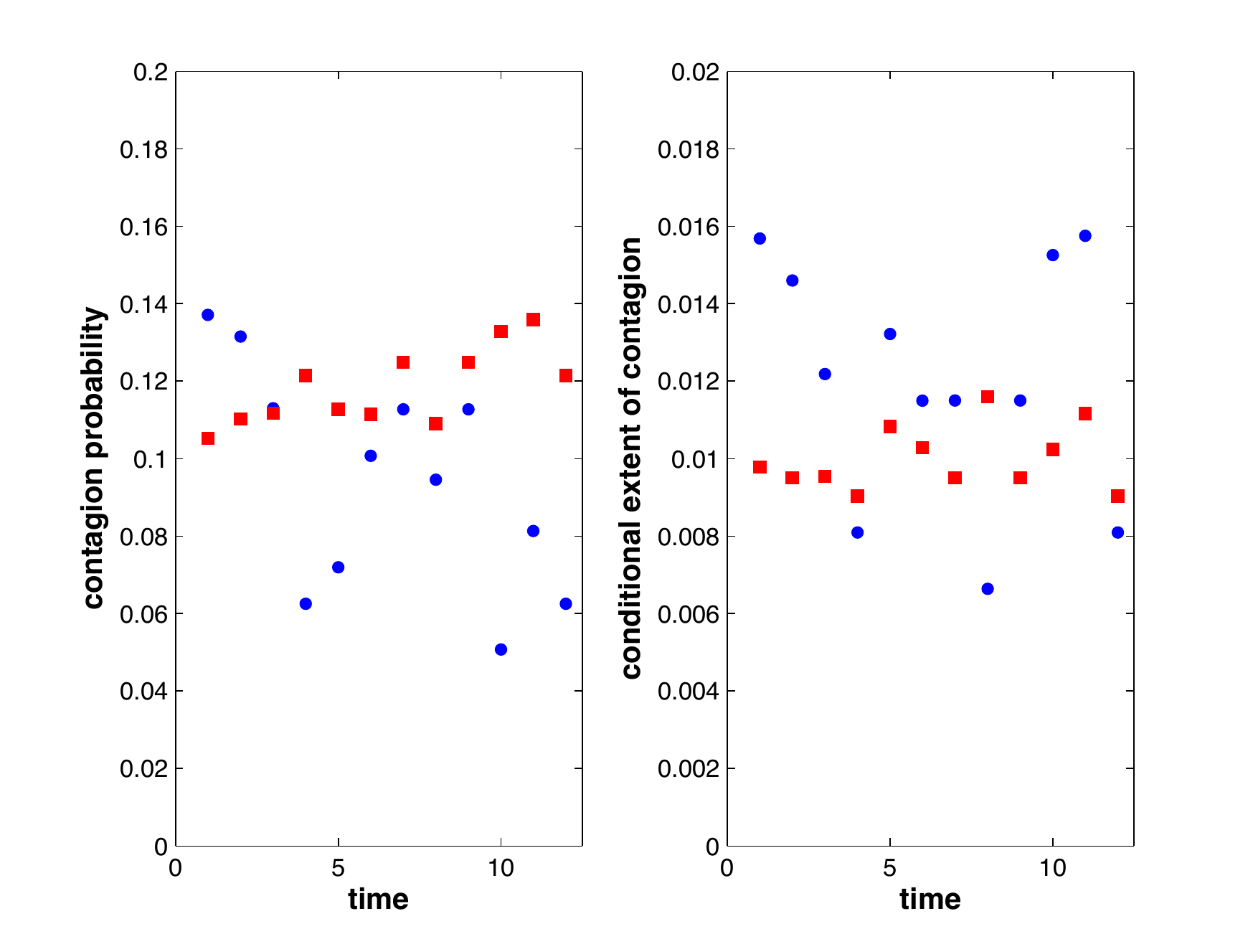}
\caption{\footnotesize {\textit{Results of stress tests. {\bf Left panel}: Contagion probability due to counterparty loss (blue dots) and roll-over risk with $f=0$ (red squares) across the different quarters of the period 2006-2008. {\bf Right panel}: Average extent of contagion due to counterparty loss (blue dots) and roll-over risk with $f=0$ (red squares) across the different quarters of the period 2006-2008. Only a small fraction of the system is affected by the initial shock.
}}}\label{figure5}
\end{center}
\end{figure}
\end{center}

In both panels, blue dots refer to contagion due to counterparty loss, while red squares refer to roll-over risk with $f=0$ (i.e., banks start hoarding as soon as they run out of liquid assets).
We see from the plot that, although the probability of contagion can be a few percent, the conditional extent of contagion is on average small both in the cases of counterparty loss and roll-over risk.

The robustness of the system does not come as a surprise, as banks have a simple way of reducing the risk associated with direct interbank exposures.
Let us consider, for instance, the case of counterparty loss. In this case a prudent bank manager will make sure that their bank is not exposed to a single other institution by more than the equity of their own bank (in some countries, like Germany, this prudential measure is actually enforced by law). It is trivial to see that if most of the banks adopt this simple measure, for which no information is required by banks other than their own balance sheet, all domino effects are damped at the start. Given this consideration, are networks of direct interbank exposures important at all? Our answer is yes, they can become important as an amplification mechanism when other contagion channels are in place. 
To make this point, in the next section we consider the case of banks with common asset holdings, and show that network effects can in certain regimes greatly amplify the effect of a sudden devaluation of the common assets.

\section{Systemic risk due to overlapping portfolios}\label{overlappingPortfolios}

We have considered so far the network of interbank claims as a channel of stress propagation in the system. Now we want to account for a different mechanism of contagion, namely the existence of overlaps between bank portfolios, as studied in \cite{Caccioli12,Huang13,Corsi13}. As a first step, we consider the effect of banks sharing a common asset in their balance sheets. For simplicity, we assume that all banks have a fraction $c$ of a common asset in their balance sheet, so that if $A_i^{tot}$ is the total assets of bank $i$, $cA_i^{tot}$ is the amount of the common asset held by bank $i$. Here $c$ is then a measure of the overlap of banks' balance sheets. 

We now ask what happens if the price of the common asset drops from a reference value $1$ to $1-\phi$. In such a situation, banks need to revise their balance sheet so that
\be
A_i^{tot}\to A_i^{tot}(1-c)+A_i^{tot} c(1-\phi)=A_i^{tot}(1-c\phi).
\ee
Depending on the level of depreciation for the asset, some banks can now go bankrupt if  their liabilities exceed the new value of their assets.\footnote{{The way overlapping portfolios are implemented here (through banks investing a share of their total assets in the same common asset) and the effect on banks of the common asset devaluation is equivalent to the introduction of a common haircut on the size of banks balance sheets, that translates into a leverage determined haircut on bank capital. We opt however for an interpretation in terms of overlapping portfolios because it would apply to more general settings as those described in \cite{Huang13,Caccioli12,Corsi13}, where the structure of overlapping portfolios is more granular}.}

\begin{center}
\begin{figure}[h]
\begin{center}
\includegraphics[width=8cm]{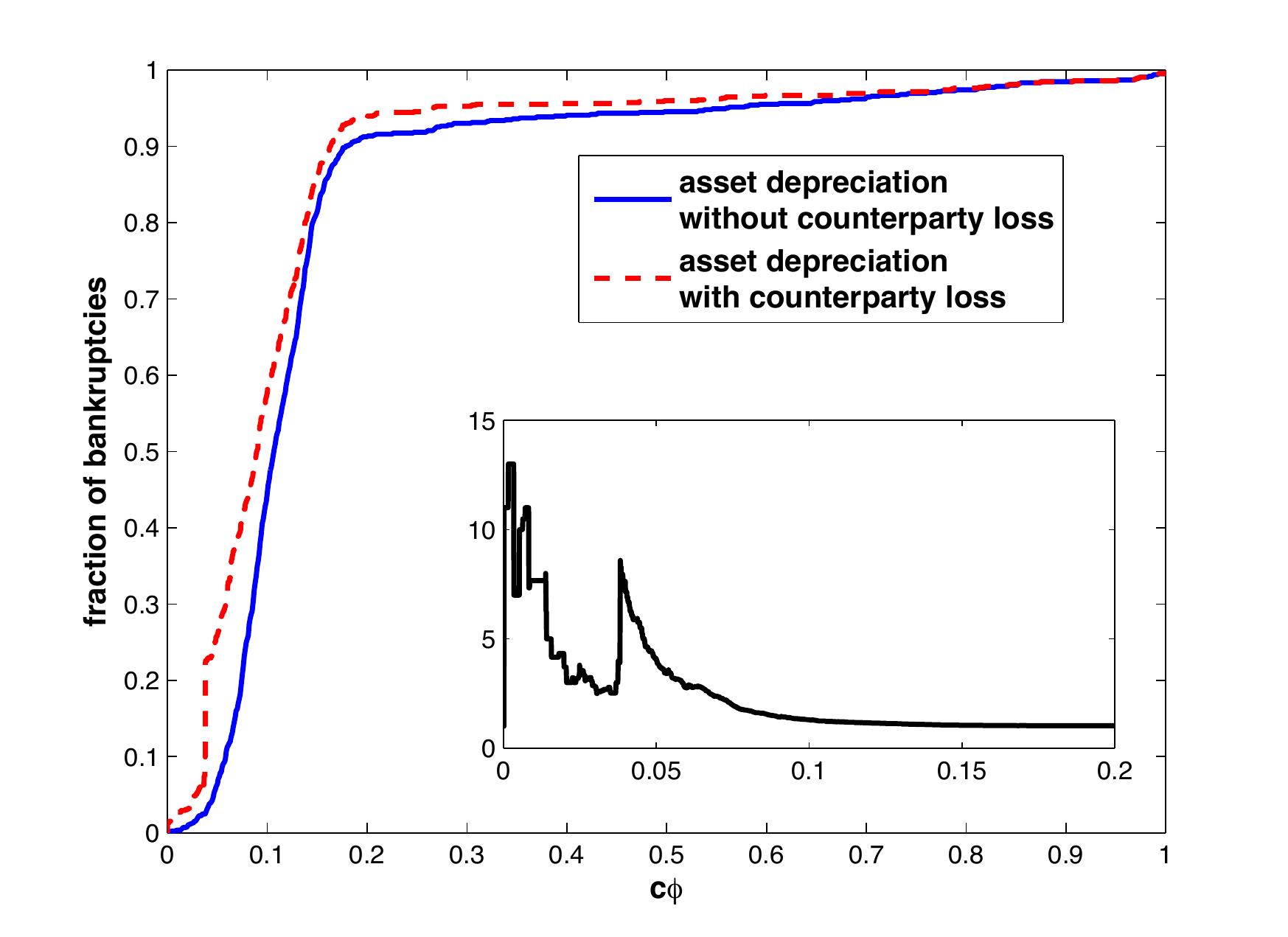}
\caption{\footnotesize {\textit{Result of stress tests. Main panel:  Fraction of bankruptcies in the system due to the depreciation of the common asset in absence (blue solid line) or presence (red dashed line) of contagion via counterparty loss. Inset: ratio of the two quantities plotted in the main panel. The network introduces a channel of contagion that can significantly increase the number of failures if $ c\phi<0.05$.}}}\label{figure6}
\end{center}
\end{figure}
\end{center}

We show in Figure \ref{figure6} the fraction of banks going underwater as a function of $c\phi$.
From the plot, we estimate a minimum value $c\phi\simeq 0.05$ resulting in $30\%$ or more of the system going down\footnote{{We note in this respect that even a small value of $c\phi$ may quickly become unrealistic. Although a value of 5\% is entirely feasible for a banking system in an adverse scenario of a multi-year stress test, earning on the remainder of banks portfolios may compensate for part of the loss.}}.
Notice that  we haven't taken into account any interaction among banks, which are independent apart from the presence of correlations in their portfolios.
We can now introduce interactions induced by the network structure, and study the effect of such interactions in combination with the common shock given by the depreciation of the common asset. The logic we used is the following: 
\begin{enumerate}
\item The common asset is depreciated.
\item Some banks go down because their liabilities exceed the new value of their assets.
\item These banks cause their creditors new losses, and contagion propagates due to counterparty loss.
\end{enumerate}

Results of simulations are shown in Figure \ref{figure6} (red dashed line). By accounting for the network structure we introduce a further channel of contagion that overall makes things worse in terms of extent of contagion. To provide a more quantitative comparison between the two cases, in the inset of the figure  we plot the ratio between the fraction of bankruptcies due to the depreciation of the common asset alone and the fraction of bankruptcies when we also account for counterparty loss. In particular, we observe that the contagion channel provided by the network can significantly enhance the number of failures if $c\phi$ is smaller than $0.05$. This amplification mechanism can be understood intuitively as follows: As we mentioned earlier the risk of default due to counterparty loss by itself can be effectively controlled by banks if they avoid being directly exposed to other institutions by an amount bigger than their own equity. This prudent measure can become ineffective, however, when the equity of banks is reduced by the devaluation of a common asset, as in this case banks may suddenly find themselves exposed to other institutions by more than their capital buffer. Therefore, contagion due to counterparty loss sets in if the devaluation of the common asset is big enough. Note that, if the common asset is highly devalued, the network of interbank exposures does not amplify contagion simply because most of the banks are already driven out of business as a result of the initial shock affecting their balance sheets. This is why the curve plotted in the inset of figure \ref{figure6} goes to one for high values of $c\phi$.

So far we have considered in this section the case of the sudden depreciation, at time $t=0$, of an asset common to all banks. This depreciation was not triggered by the dynamics of the system, but was due to an exogenous shock.
We now consider what happens when asset devaluations are instead endogenously induced by banks that liquidate their portfolios. Let us consider the following stress protocol:
\begin{enumerate}
\item A bank $q$ is initially selected for bankruptcy.
\item When a bank $i$ goes bankrupt, its interbank liabilities are not repaid and its portfolio of illiquid assets is liquidated.
\item The liquidation process induces a devaluation of the common asset proportional to the relative size of bank $i$, i.e.,  ${A_i^{tot}}/{\sum_j A_j^{tot}}$\footnote{{The assumption that the drop in price associated with a bank liquidating its position on the asset is proportional to the relative size of the bank with respect to the entire banking system may be severe as a significant fraction of the asset may be held outside the banking system.}}.
\item All banks suffer a loss that is due to the devaluation of the common asset. Banks with direct interbank exposures to $i$ take a further hit due to counterparty loss.
\item If new banks are bankrupt, return to Step 2. Exit otherwise.
\end{enumerate}

\begin{center}
\begin{figure}[h]
\begin{center}
\includegraphics[width=8cm]{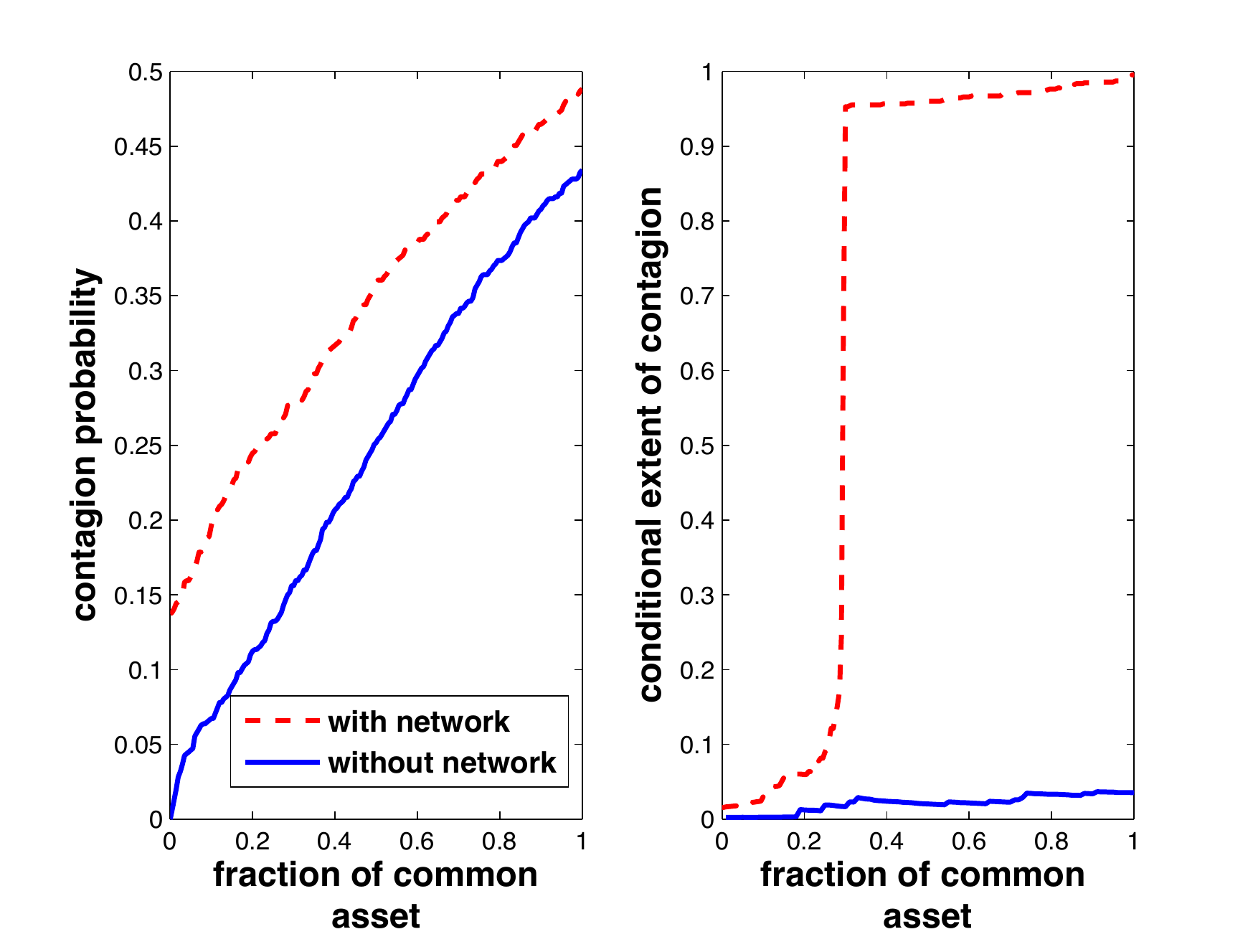}
\caption{\footnotesize {\textit{Results of stress tests. {\bf Left panel}:  Contagion probability due to liquidation of the common asset in absence (blue solid line) or presence (red dashed line) of contagion due to counterparty loss. {\bf Right panel}: Conditional extent of contagion due to liquidation of the common asset in absence (blue solid line) or presence (red dashed line) of contagion due to counterparty loss. The presence of a network of direct interbank exposures can substantially amplify contagion due to liquidation of common assets.}}}\label{figure7}
\end{center}
\end{figure}
\end{center}

In Figure \ref{figure7} we plot the probability and average extent of contagion measured for the first quarter of 2006 as a function of $c$ (fraction of total assets invested by each bank in the common asset). Each panel shows two lines. Blue solid lines refer to the case in which the network of direct interbank exposures is not accounted for (i.e.,  we ignore losses due to defaulted counterparties). In contrast, red dashed lines show what happens when interbank exposures enter in the picture.  Clearly, the presence of the network increases both the probability and extent of contagion. The contagion probability gradually increases from about $0.14$ to $0.49$ as $c$ increases from $0$ to $1$. More interestingly, we observe that the interaction between contagion due to counterparty loss and contagion due to common asset holding induces a sudden jump in the average extent of contagion for $c\simeq 0.3$. 
A comment is in order in this respect. The jump observed in the average extent of contagion is not due to the fact that around $c\simeq 0.3$ large cascades of bankruptcies suddenly become possible because of contagion due to counterparty loss. In fact, large cascades can be observed also if counterparty loss is not accounted for. Without counterparty loss no jump is observed because the probability of observing  a large cascade is always small (the probability of having more than 10 bankruptcies is for instance less 2\%). If counterparty loss is added to the picture, instead, the probability of observing a large cascade becomes significant around $c\simeq 0.3$ (the probability of having more than 10 bankruptcies is for instance more than 30\%).

\section{Conclusion}\label{conclusions}
This paper focused on assessing the importance of networks of direct interbank exposures for financial contagion under different stress scenarios. Recent empirical and theoretical work suggests that, in real interbank networks, direct interbank exposures by themselves do not contribute significantly to financial contagion. The point we address here was that of understanding whether contagion due to direct exposures can nonetheless become relevant in interaction with other contagion mechanisms, and we find that contagion due to counterparty loss can amplify the stress induced by the presence of common asset holdings in bank balance sheets.

In this paper we considered data concerning the Austrian banking system in the period 2006-2008. In the first part of the paper we provided a statistical characterization of the system, in particular for its balance sheet and network properties. We found that there is a power law regime in the distribution of banks' size and that the network of interbank exposures is characterized by a heavy-tailed distribution of node degree and by negative correlations between degrees of neighboring nodes. Such properties have been shown to be relevant in terms of financial contagion and have been observed in the past for other data-sets of national banking systems.
This fact suggests that the results of stress tests presented in this paper do not apply only to the specific data here considered.

We performed stress tests of the Austrian banking system according to different stress protocols, and using different contagion mechanisms: counterparty loss, roll-over risk and devaluation of a common asset. 
The system is shown to be fairly stable if counterparty loss is the only contagion mechanism in place. Roll-over risk is also shown not to play a major role by itself.  

The network of direct interbank exposures becomes an important mechanism of stress amplification, however, when it interacts with contagion due to overlapping portfolios. If a fraction of bank total assets is invested in a common asset, we showed that contagion due to counterparty loss can greatly amplify the fraction of bankruptcies occurring in the system after a macroeconomic shock leading to the devaluation of the common asset. Moreover, we showed that contagion due to counterparty loss induces a discontinuity in the average number of bankruptcies when the asset devaluations due to bankrupted banks liquidating their portfolios is also accounted for.

\appendix
\section{Appendix: Comparison with null models: topological properties}

We have mentioned in Section \ref{data} of the main text that fixing the degree sequence of the interbank network is enough to explain simple topological properties like degree correlations and local clustering.

\begin{center}
\begin{figure}[h]
\begin{center}
\includegraphics[width=3.5cm]{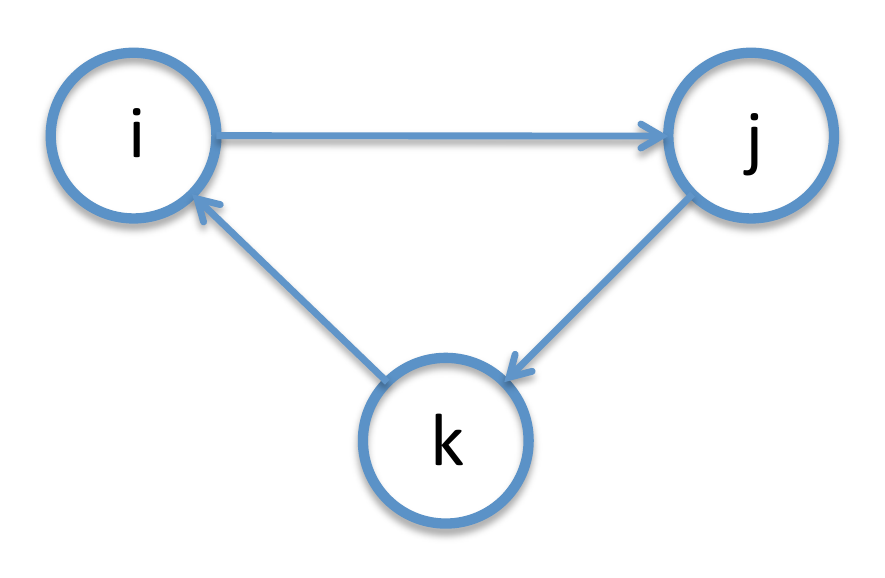}
\includegraphics[width=3.5cm]{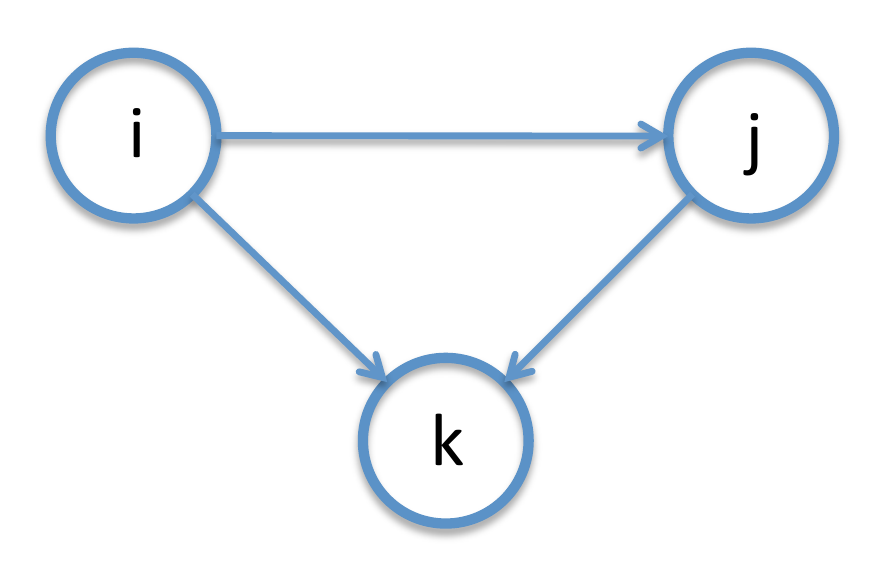}
\caption{\footnotesize {\textit{These are the two possible 3-clique motifs on 3 nodes in a directed network: On the left is a directed cycle, on the right, not a cycle and node $i$ serves as a ``source" and node $k$ serves as a ``sink." }}}\label{toycyclefigs}
\end{center}
\end{figure}
\end{center}

\begin{center}
\begin{figure}[h]
\begin{center}
\includegraphics[width=8cm]{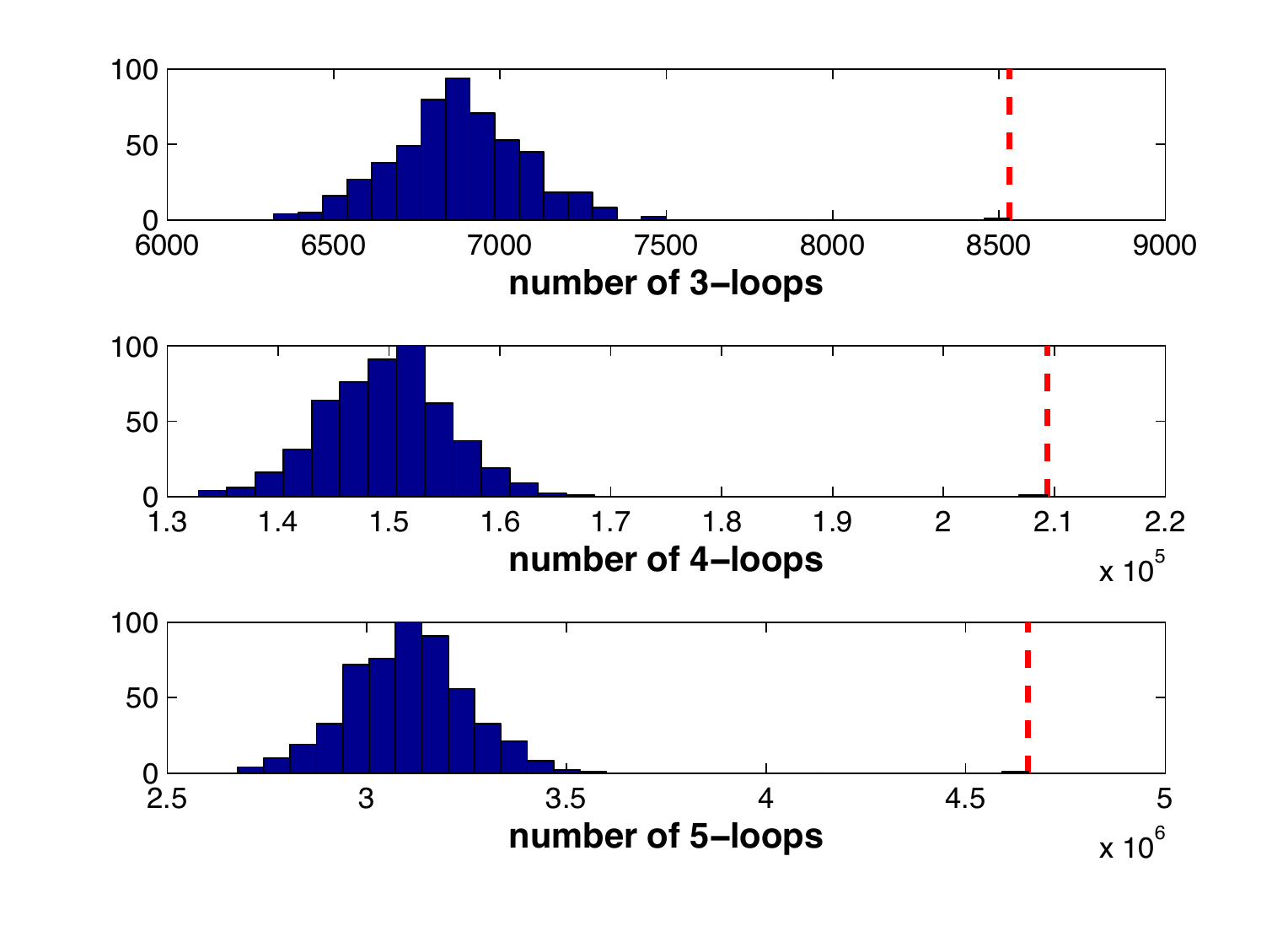}
\caption{\footnotesize {\textit{Directed loops. Histograms of number of directed loops of order $3$ (upper panel), $4$ (middle panel) and $5$ (lower panel) obtained from $10^3$ directed networks with same in-degree and out-degree distributions of the real network corresponding to the first quarter of 2006. The red dashed vertical lines correspond to the values measured for the real network. The density of these local patterns is much higher in the real network than in the random ones (the difference in the three panels is higher than 8 standard deviations). This suggest that these motifs might serve a purpose specific to the nature of the interbank networks.}}}\label{loops}
\end{center}
\end{figure}
\end{center}

The connectivity encoded in the local clustering is interesting and relevant to the networks, but in addition we should also account for the directionality in these triangles. Directionality encodes the actual lending patterns and not just the dependence. A directed cycle $i \longrightarrow j \longrightarrow k  \longrightarrow i$ indicates a kind of liability pattern quite different in nature from $i \longrightarrow j \longrightarrow k \mbox{ and } i  \longrightarrow k$ (see Figure~\ref{toycyclefigs}). In the network literature such connectivity patterns are often called {\em motifs} and are highly relevant in the understanding of social and biological networks \cite{Milo02}. Considering all quarters in the data-set, in reference to the motifs shown in Figure~\ref{toycyclefigs}, we found an average of $8\times 10^3$ directed cycles and $1.4\times 10^5$ cycles of length $3$ with one source and one sink.

We report in Figure \ref{loops} a comparison between the number of directed cycles of length $3$, $4$ and $5$ measured in the network corresponding to the first quarter of 2006, and the distribution obtained for the same quantity for random networks with same in-degree and out-degree sequences.

Given the relatively high number of short cycles that have been measured in the real network (Figure \ref{loops}),  it would be interesting to speculate whether these local patterns serve a purpose i.e. specific to the nature of interbank lending. In order to test this hypothesis, similar analysis should be carried on on different interbank networks.

\section{Appendix: Comparison with null model: stress tests}
In this section we compare results of stress tests obtained from real and synthetic networks.
We start by looking at counterparty loss, and we focus on the first quarter of $2006$. The null model is the ensemble of networks obtained by randomly rewiring the links of the real network, that is the random ensemble of networks that has the same in-degree and out-degree sequences of the real network.

An observation concerning the rewiring procedure is in order at this time. It is now important to remember that we are dealing with a weighted directed network, where each entry of the liability matrix represents a loan from one institution to another. During the rewiring procedure we have to decide what to do with the weight attached to each link that is being rewired. As far as counterparty loss is considered, given that contagion comes from the asset side of the balance sheet,  the weight of each rewired link stays with the bank for which that specific link represents an asset. Total liabilities 
are then adjusted at the end of the rewiring process in such a way to match the real equity of institutions.

\begin{center}
\begin{figure}[h]
\begin{center}
\includegraphics[width=8cm]{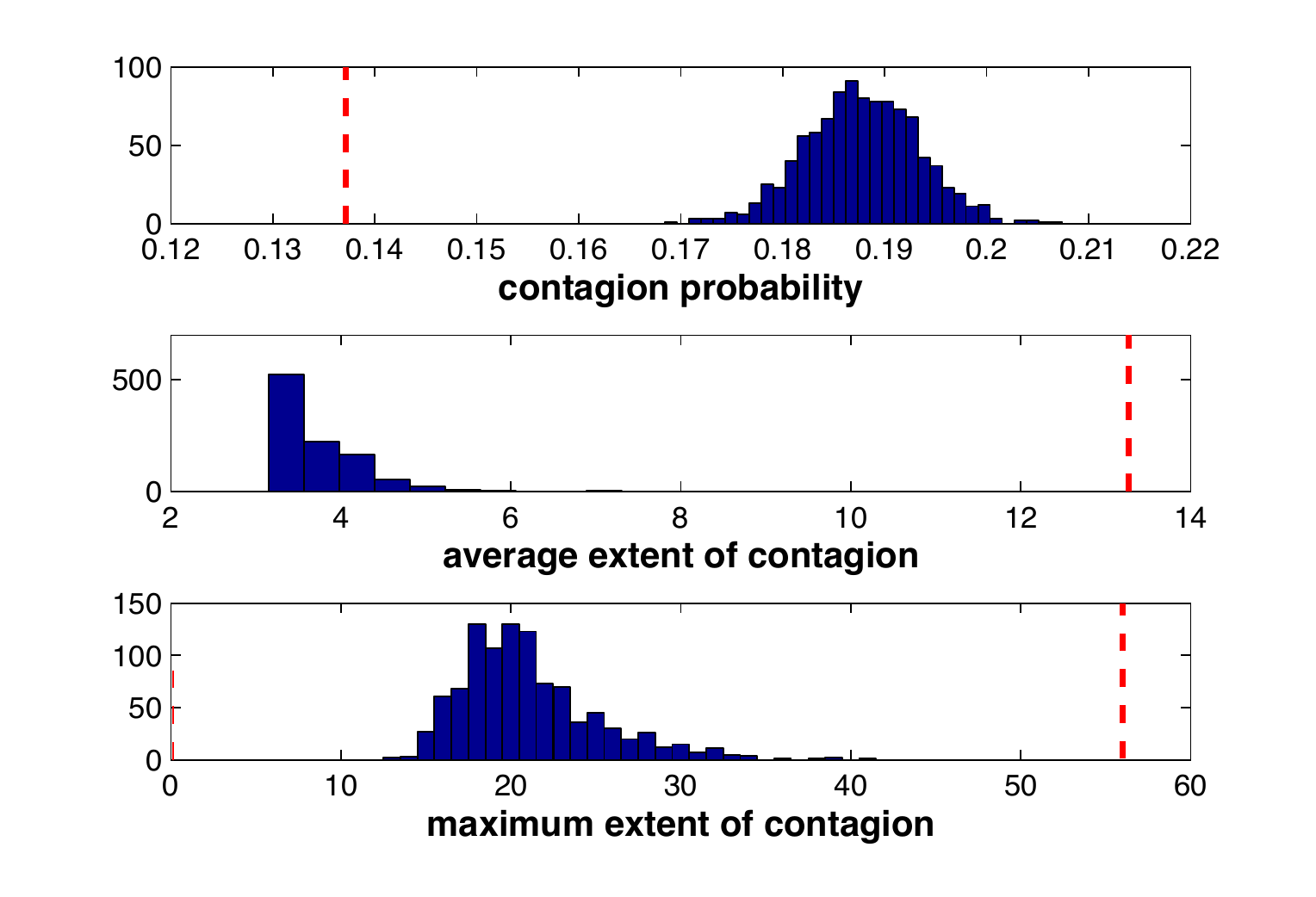}
\caption{\footnotesize {\textit{ Stress tests, comparison with null model. Counterparty loss in the first quarter of 2006: histograms of contagion probability (top panel), average extent of contagion (middle panel) and maximum extent of contagion (bottom panel) obtained from $10^3$ random networks with same degree sequence as the real one. The red dashed vertical lines are the corresponding quantities measured for the real network. Synthetic data overestimate the probability of contagion, while average and maximum extent of contagion are correctly predicted. 
}}}\label{contprob}
\end{center}
\end{figure}
\end{center}

Results for the first quarter \footnote{ Similar results have been obtained for the other quarters of 2006 as well as 2007 and 2008.}  of 2006 are reported in Figure \ref{contprob} for three different quantities: 
\begin{itemize}
\item [i)]
contagion probability -- fraction of nodes whose initial failure triggers {\bf at least} one more failure in the system; 
\item [ii)] {\bf conditional} average extent of contagion -- average number of banks going down {\bf if} contagion takes place; 
\item [ iii)] maximum extent of contagion -- maximum number of bankruptcies measured for a network.
\end{itemize}

First of all, we observe in the upper panel of Figure \ref{contprob} that the networks generated with the null model we considered overestimate the contagion probability measured for the real network. Given the fact that we define a contagion event as an event where at least one bank goes bankrupt as a consequence of the initial perturbation, and that the exposures associated with the links are preserved in the rewiring procedure, the discrepancy between contagion probability in real and synthetic networks must be explained in terms of a different arrangement of  ``critical'' links, where by critical link we mean an interbank exposure that is enough to trigger the failure of one bank should its counterparty go bankrupt. Indeed, the smaller contagion probability observed for the real network is due to the fact that the same number of critical links is spread between fewer nodes than for synthetic networks.
 
Figure \ref{contprob} also shows that synthetic networks largely underestimate average and maximum extent of contagion. A possible explanation could be in the different local structure of real and synthetic networks. 

We now turn to the case of roll-over risk.
Notice that now, given that contagion propagates through the funding channel,  the weight of each rewired link stays with the bank for which that specific link represents a liability (i.e., a source of funding). Total assets 
are then adjusted at the end of the rewiring process in such a way to match the real equity of institutions.

\begin{center}
\begin{figure}[h]
\begin{center}
\includegraphics[width=8cm]{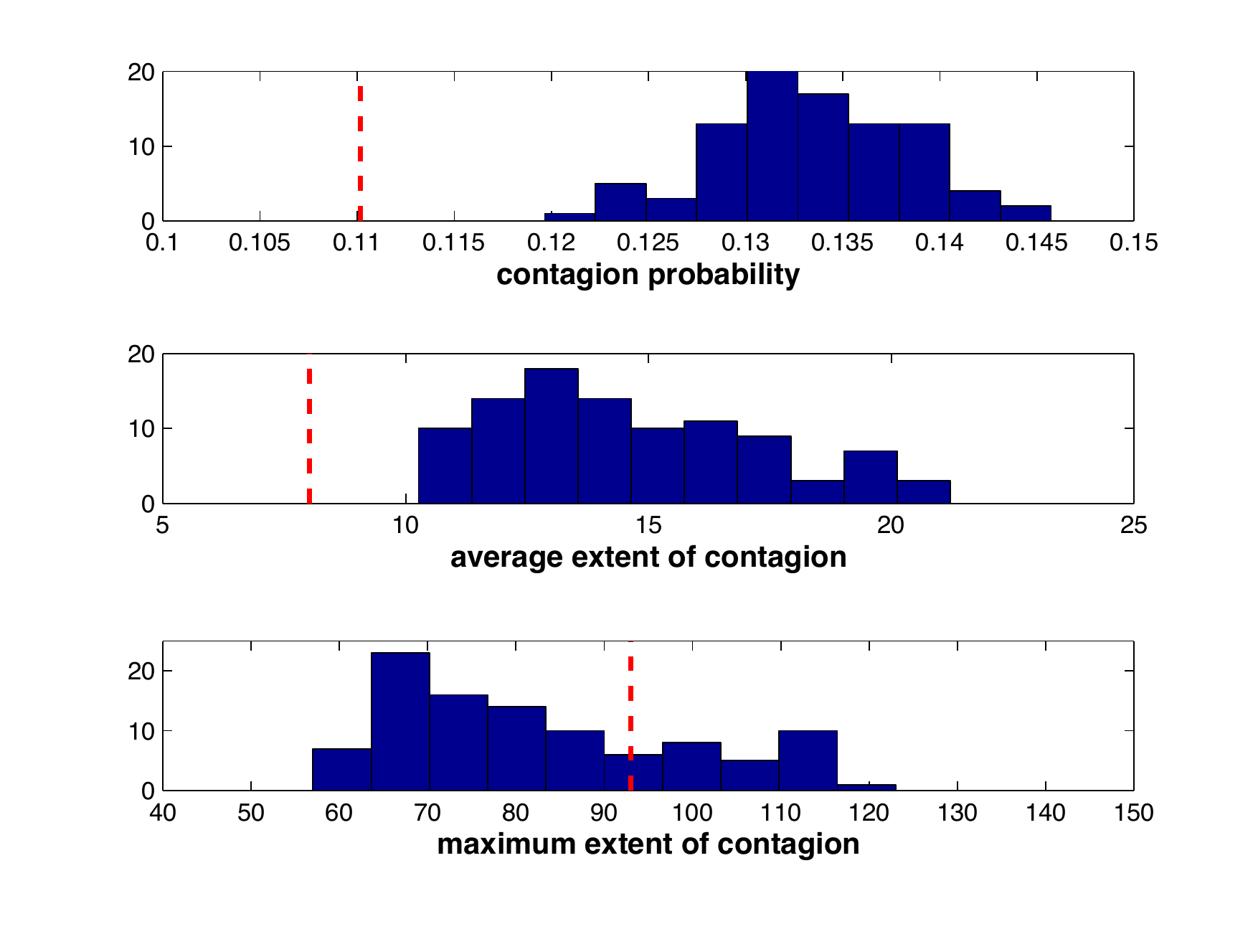}
\caption{\footnotesize {\textit{Stress tests, comparison with null model. Rollover risk in the second quarter of 2006: histogram of contagion probability (top panel), average extent of contagion (middle panel) and maximum extent of contagion (bottom panel) obtained from $10^3$ random networks with same degree sequence as the real one. The red dashed vertical lines are the corresponding quantities measured for the real network. Synthetic data overestimate the probability of contagion, as well as  the average extent of contagion. 
}}}\label{hoarding2}
\end{center}
\end{figure}
\end{center}

In Figure \ref{hoarding2}, we report results obtained for the second quarter of 2006 as an example of the pattern observed for the other periods. In this case, the null model overestimates the probability of contagion and the average extent of contagion. As before, vis-a-vis the contagion probability, the discrepancy is explained in terms of a different arrangement of critical links among nodes.  

The comparison of real data with the null model considered here makes it clear that knowledge of the degree sequence is not enough to correctly estimate probabilities and extent of contagion due to counterparty loss or liquidity hoarding. In particular, we considered for both channels of contagion a random ensemble of networks that only preserved the actual network's   degree sequence and number of nodes. Such constraints are strong enough to force the network to build strong negative degree correlations and to induce a high average clustering, but this is not enough to account for the properties observed for contagion due to funding or counterparty loss.  At least part of the discrepancies observed between the real system and the null model can been explained in terms of a different arrangement of critical links.

\section*{Acknowledgments}
This work was supported by the National Science Foundation under grant 0965673,
by the European Union Seventh Framework Programme FP7/2007-2013 under
grant agreement CRISIS-ICT-2011-288501 and by the Sloan Foundation. The authors would like to thank Martin Summer and Claus Puhr for their help in sharing the data and for useful discussions. We also warmly thank Stefan Thurner for useful discussions. D. R. acknowledges the support of the Dartmouth
College Neukom Institute for Computational Science.
\bibliographystyle{plain}
\bibliography{AB_refs}

 \end{document}